\documentclass[twocolumn,prd,amsmath,superscriptaddress]{revtex4}
\usepackage{graphicx}
\usepackage{latexsym}
\usepackage{amsmath}
\usepackage{amssymb}
\usepackage{graphics}
\usepackage{color}
\usepackage{hyperref}
\usepackage{bm}

\newcommand{\calA}{$\mathcal A$}
\newcommand{\calB}{$\mathcal B$}
\newcommand{\dVeff}{V_{\mathrm{eff},\phi}}

\newcommand{\Ekin}{E_\mathrm{kin,SN}}
\newcommand{\Emono}{E_\mathrm{mono,SN}}
\newcommand{\epsenv}{\epsilon_\mathrm{E}}
\newcommand{\fRo}{f_{\mathrm{R}0}}
\newcommand{\Lsun}{L_\odot}
\newcommand{\Mc}{M_\mathrm{C}}
\newcommand{\Me}{M_\mathrm{E}}
\newcommand{\meff}{m_\mathrm{eff}}
\newcommand{\Mpl}{M_\mathrm{Pl}} 
\newcommand{\Msun}{M_\odot}
\newcommand{\phiB}{\phi_\mathrm{B}}
\newcommand{\phimax}{\phi_\mathrm{max}}
\newcommand{\phimin}{\phi_\mathrm{min}}
\newcommand{\Pmono}{P_\mathrm{mono}}
\newcommand{\Psie}{\Psi_\mathrm{env}}
\newcommand{\Psis}{\Psi_\mathrm{s}}
\newcommand{\rc}{r_\mathrm{C}}
\newcommand{\re}{r_\mathrm{E}}
\newcommand{\rhoc}{\rho_\mathrm{C}}
\newcommand{\rhoe}{\rho_\mathrm{E}}
\newcommand{\rhomo}{\rho_{\mathrm{m}0}}
\newcommand{\rsun}{r_\odot}
\newcommand{\RJ}{R_\mathrm{J}}
\newcommand{\RJo}{R_{\mathrm{J}0}}

\newcommand{\ve}{v_\mathrm{E}}
\newcommand{\Veff}{V_\mathrm{eff}} 

\begin{document}

\title{Monopole radiation in modified gravity}
\author{Amol Upadhye}
\affiliation{High Energy Physics Division, Argonne National Laboratory, 9700 S. Cass Ave., Argonne, IL 60439}%
\author{Jason H.~Steffen}
\affiliation{Northwestern University, 2131 Tech Drive, Evanston, IL 60208}%
\date{\today}

\begin{abstract}
Modifications to General Relativity typically introduce a scalar degree of
freedom.  Experimental constraints require that the ``fifth'' force produced
by this field must be screened in high density environments.  An important
consequence of the screening mechanism is that the scalar charge of a
spherically symmetric expanding or pulsating object, such as a supernova or
variable star, is not conserved and monopole radiation can be emitted thereby.
We calculate the energy loss rate due to scalar monopole radiation in supernovae and variable stars for a generalized model of $f(R)$ gravity in which the matter coupling strength $\beta$ is allowed to be much larger than unity.  For  models which become screened at the gravitational potential of our galaxy, monopole radiation constraints require $\beta \lesssim 100$.  
\end{abstract}

\maketitle

\section{Introduction}
\label{sec:introduction}

Fifteen years after the discovery of the cosmic acceleration, its cause remains the greatest mystery in cosmology.  Since the latest data~\cite{Ade_etal_2013xvi} are consistent with a constant vacuum energy density, attention is turning to modified gravity models~\cite{Jain_Khoury_2010}, which are described at low energies by scalar fields coupled to matter.  
Fifth forces mediated by such scalars are tightly constrained in the laboratory and the solar system, so the only viable modified gravity models are those which screen fifth forces at high densities.  
Three classes of models achieve
this: chameleon models become massive at high densities~\cite{Khoury_Weltman_2004a,Khoury_Weltman_2004b,Brax_etal_2004}, symmetrons decouple from matter in a symmetric phase~\cite{Olive_Pospelov_2007,Hinterbichler_Khoury_2010}, and Galileons suppress their gradients through the Vainshtein mechanism~\cite{Vainshtein_1972,Nicolis_Rattazzi_Trincherini_2008}.

$f(R)$ modified-action gravity models, which add to the Ricci scalar $R$ some function $f(R)$ in the Einstein-Hilbert action, employ chameleon screening.  The effective scalar field outside an object is sourced only by a thin shell of matter near the object's surface, whose thickness depends on the density and geometry of the object.  Inside a screened environment such as a galaxy cluster, stars which would be unscreened in the vacuum may become screened.  In an isolated dwarf galaxy without such {\em environmental screening,} the outer envelopes of stars may remain unscreened, leading to residual fifth forces~\cite{Cabre_etal_2012}.  Thus $f(R)$ gravities predict an increase in the luminosities of such stars, and, hence, unscreened dwarf galaxies~\cite{Davis_etal_2012}.  The period-luminosity relation of variable stars is also modified in unscreened regions, providing the leading constraint on $f(R)$ models~\cite{Jain_Vikram_Sakstein_2012}.  

Geometry-dependent screening implies that the effective scalar charge of an object whose size changes with time is not conserved.  Thus the spherical pulsation of a variable star, or the sudden explosion of a supernova, will lead to the emission of scalar monopole radiation.  Since monopole radiation could affect the constraints of~\cite{Jain_Vikram_Sakstein_2012}, understanding it is important.  Here we extend the linearized analysis of~\cite{Silvestri_2011}, solving the fully nonlinear chameleon equation in a screened star for generalized $f(R)$ modified gravity models.  We use our solution to compute the energy loss due to monopole radiation in variable stars and core-collapse supernovae.  Since chameleon screening makes the scalar field insensitive to the details of stellar structure, we use a simple stellar model in which a core of constant size and density is surrounded by a uniform envelope whose radius varies with time. Such a model is adequate for estimating the effects of monopole radiation.  We find that, in $f(R)$ gravity, the energy lost through scalar monopole radiation is too small by several orders of magnitude to affect current constraints.  In generalized $f(R)$ models which allow stronger matter couplings, monopole emission from supernovae could substantially slow down the ejected matter in the aftermath of the explosion.  We use this predicted deceleration to estimate constraints on generalized $f(R)$ models which would result from the observation of a supernova in an unscreened region.

This article is organized as follows.  Section~\ref{sec:monopole_radiation} describes our generalized $f(R)$ gravity and our stellar models, then predicts the energy lost to monopole radiation in variable stars and supernovae.  Constraints are computed and discussed in Sec.~\ref{sec:results_and_discussion}.

\section{Monopole Radiation}
\label{sec:monopole_radiation}

{\bf{\em Modified gravity in the Einstein frame:}}
$f(R)$ gravity adds to the Jordan-frame Einstein-Hilbert action $\RJ$ a function $f(\RJ)$.  In currently allowed models $f(\RJ)$ is close to a constant.  For example, in the Hu-Sawicki model~\cite{Hu_Sawicki_2007,Hu_Sawicki_2007b} $f(\RJ) = -2\Lambda/ \left( 1 + \fRo \RJo^{\ell+1}/(2\ell\Lambda\RJ^\ell) \right)$, $|\fRo|$ is constrained to be much smaller than unity; $f(\RJ)$ is well-approximated by $-2\Lambda$ plus a power law in $\RJ$.  Conformally transforming to the Einstein frame, in which the form of the Einstein-Hilbert action is recovered, we find a canonically-normalized scalar field $\phi$ with a potential 
$V(\phi)$
and a matter interaction $-\beta \phi T_\mu^\mu / \Mpl$, where $\beta=1/\sqrt{6}$ and $T_{\mu\nu}$ is the stress tensor of ordinary matter.  A power-law term $\propto \RJ^{-\ell}$ in $f(\RJ)$ implies a power-law term $\propto |\phi|^n$ in $V$, where $n = \ell/(\ell+1)$~\cite{Jain_Khoury_2010}.  We are primarily concerned with $\ell > 0$, that is, $0 < n < 1$.  Such models have the largest field variations, maximizing monopole radiation, and are inaccessible to laboratory experiments~\cite{Upadhye_2012}.

Consider a scalar with $V(\phi) = V_0 + -\gamma M_\Lambda^{4-n} \phi^n$ and effective potential $\Veff = V(\phi) - \beta \phi T_\mu^\mu / \Mpl$, where $0 < n < 1$ and $M_\Lambda = 2.4\times 10^{-3}$~eV is the dark energy scale.  In bulk nonrelativistic matter of density $\rho \approx -T^\mu_\mu$, the bulk field value $\phiB$ is given by $\dVeff(\phiB(\rho))=0$, and the effective mass $\meff(\phi) = V_{,\phi\phi}^{1/2}$ where the subscript $,\phi\phi$ denotes two derivatives with respect to $\phi$.  Known as the ``chameleon effect,'' the increase of $\meff$ with $\rho$ allows the field to evade local fifth force constraints~\cite{Khoury_Weltman_2004a,Khoury_Weltman_2004b,Brax_etal_2004}.  

Quantum corrections from matter loops do not change the form of this matter coupling, but they are expected to change the value of $\beta$~\cite{Hui_Nicolis_2010}.  Thus we generalize $f(R)$ gravity to allow arbitrary $\beta$; this amounts to the addition of a scalar field kinetic term in the Jordan frame. Scalar loop corrections may lead to large modifications to $V(\phi)$ at stellar densities, where $\phi$ differs by many orders of magnitude from its value in the cosmic background~\cite{Upadhye_Hu_Khoury_2012}.  However, we will see that monopole radiation is dominated by field fluctuations near the cosmic background value.  Quantum corrections to the field profile in the stellar interior will not affect our results.

Consider an object of size $\rc$ and density $\rhoc$ in the cosmological background density $\rhomo \approx 2.5\times 10^{-30}$~g/cm$^3$.  The static field equation $\nabla^2 \phi = V_\phi + \beta \rho / \Mpl$, with $V_\phi = dV/d\phi$, simplifies in the limit of small $\rc$.  Since the perturbation $\delta\phi = \phi - \phiB(\rhomo)$ is small, the potential term is negligible, and the resulting equation looks like the Poisson equation for the gravitational field $\Psi$.  In this linear regime, $\delta\phi = 2\beta \Mpl \Psi$. For objects with large $\rc$, the perturbation is large enough that $V_\phi$ in the field equation is no longer negligible; it takes a large negative value which partially cancels (``screens'') the density term.  Since $\phi>0$, linearity must break down for $\delta\phi < -\phiB(\rhomo)$.  This corresponds to a gravitational potential $|\Psi| = \Psis$, where the ``self-screening parameter'' $\Psis = \phiB(\rhomo) / (2\beta\Mpl)$ is determined by the current cosmic background density.  For convenience we treat $\Psis$ as a parameter of the potential, instead of $\gamma=\beta\rhomo(n\Mpl M_\Lambda^3)^{-1} (2\beta\Mpl\Psis/M_\Lambda)^{1-n}$:
\begin{eqnarray}
V(\phi)
&=&
V_0 - \frac{\beta\rhomo}{n\Mpl}(2\beta\Mpl\Psis)^{1-n}\phi^n
\\
\phiB(\rho)
&=&
2\beta\Mpl \Psis
(\rhomo/\rho)^\frac{1}{1-n}
\\
\meff(\phi)^2
&=&
\frac{\rhomo(1-n)}{2\Mpl^2\Psis} 
\left(\frac{2\beta\Mpl\Psis}{\phi}\right)^{2-n}.
\end{eqnarray}

{\bf{\em Variable stars and supernovae:}}
As an approximate model of a variable star we assume a core of constant density $\rhoc$ and radius $\rc$ surrounded by an envelope of density $\rhoe$ and radius $\re$.  The envelope contains a fraction $\epsenv$ of the total mass $M_\star$.  Pulsation of the star can be described by a time-variation of $\re$ at constant $\rhoc$, $\rc$, $\epsenv$, and $M_\star$, with $\Delta r \equiv \re(\max)-\re(\min)$.  For concreteness we consider two stars, \calA~and \calB, shown in Table~\ref{t:stars}, which roughly correspond to a dwarf Cepheid and a classical Cepheid, respectively.  
The envelope mass fractions in the table are estimated using energy conservation.  Matter in the envelope is pushed outward by radiation pressure, so the total change $\Delta U$ in the star's gravitational potential energy must be smaller than the radiated energy $L\tau$.  We estimate $\epsenv$ by assuming $\Delta U \sim 0.1 L\tau$ with luminosities $L_\mathcal{A} \sim 10 L_\odot$ and $L_\mathcal{B} \sim 10^4 L_\odot$.  The gravitational potentials of these envelopes considered independently of their cores are $-G M_\star \epsenv / rE(\min) = -1.7 \times 10^{-16}$ for star \calA~and $-2.2\times 10^{-13}$ for star \calB.  As we will see, the emitted monopole power can be divided into three regimes: at low $\Psis$, the core and envelope will be screened; at intermediate $\Psis$, the envelope will be unscreened; and at large $\Psis$, the entire star will be unscreened, and monopole radiation will drop rapidly.

\begin{table}[tb]
\begin{center}
\begin{tabular}{|l|cccccc|}
\hline
       & $M_\star$ & $\epsenv$ & $\rc$ & $\bar{\re}$ & $\Delta r$ & $\tau$ \\
\hline
star \calA~ & 2   & $10^{-10}$ & $2\rsun$ & $3\rsun$ & $1\rsun$ & $10^4$~s\\
star \calB~ & 10  & $10^{-6}$  & $90\rsun$ & $100\rsun$ & $10\rsun$ & $10^6$~s\\
\hline
\end{tabular}
\end{center}
\caption{Two models for variable stars.  Listed are the total mass $M_\star$ of each star, the mass fraction $\epsenv$ in the envelope, the core radius $\rc$, the mean envelope radius ${\bar \re} = (\re(\min)+\re(\max))/2$, the variation in radius $\Delta r = \re(\max) - \re(\min)$, and the oscillation period $\tau$.  \label{t:stars}}
\end{table}

\begin{figure}[tb]
\begin{center}
\includegraphics[angle=0,width=3.3in]{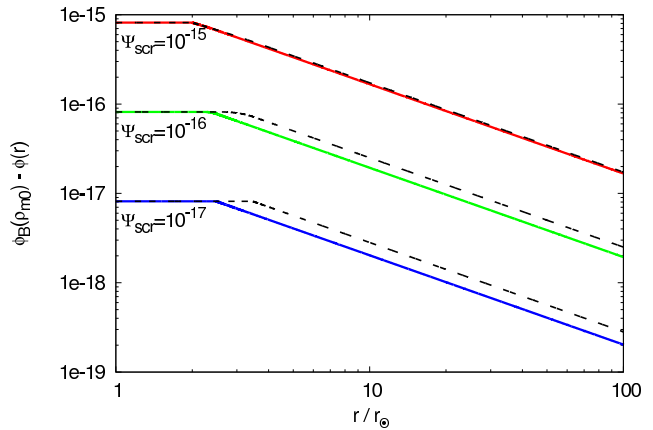}
\caption{$\phi(r)$ for a variable star.  Solid colored lines show the field for $\re = 2.5\rsun$, while dashed lines show $\re = 3.5 \rsun$.  Scalar potentials with $n=1/2$ and several $\Psis$ are shown. \label{f:phiD}}
\end{center}
\end{figure}

The matter in the stellar envelope is moving at nonrelativistic speeds.  Therefore, at the turning points of the oscillation, $\re(\min)$ and $\re(\max)$, the field $\phi(r,t)$ relaxes to its static profiles $\phimin(r)$ and $\phimax(r)$ solving $\nabla^2 \phi = \dVeff$.  
We compute $\phi(r)$ for these two $\re$ by integrating this static field equation for given $n$ and $\Psis$, assuming a bulk density $\rhomo$.  Figure~\ref{f:phiD} shows the field for star \calA, for models in which the stellar envelope is screened ($\Psis = 10^{-17}$), marginally screened ($\Psis = 10^{-16}$), and unscreened ($\Psis = 10^{-15}$); in all cases, the stellar core is screened.  In the screened cases, $\phi \ll \phiB(\rhomo) = 2\beta\Mpl\Psis$ right up to the outer edge of the star, $r\approx \re$, after which $2\beta\Mpl\Psis - \phi(r)$ falls of as $r^{-1}$.  Thus $r (\phimax(r) - \phimin(r)) \approx 2\beta\Mpl\Psis \Delta r$.  We have verified that $r(\phimax-\phimin)$ is indeed constant by integrating out to $r = 10^9 \rsun$.

\begin{figure}[tb]
\begin{center}
\includegraphics[angle=270,width=3.3in]{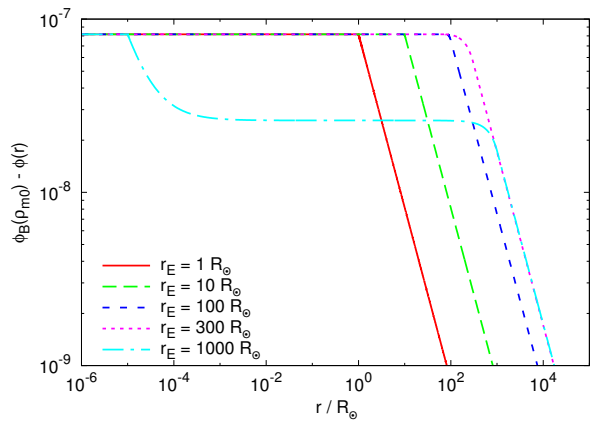}
\caption{$\phi(r)$ in a simple model of a core-collapse supernova, for several values of the envelope radius $\re$, in a modified gravity model with $n=1/2$ and $\Psis = 10^{-7}$.  The field profile changes with $\re$ until the linear regime, $\re \sim 200 \rsun$.  \label{f:phiSN}}
\end{center}
\end{figure}

This core-and-envelope model can also be applied to a core-collapse supernova.  In the aftermath of the supernova explosion, a neutron star core of mass $\Mc \sim \Msun$ and size $\rc \sim 10^{-5}\rsun$ remains at the center, while an envelope of mass $\Me = 10~\Msun$ expands outward at a rate $\ve \equiv d\re / dt \sim 10^4$~km/s.  In our model, the constant-density envelope undergoes homologous expansion, with velocity $v(r) \propto r$, which approximates the $1$-D numerical calculations of~\cite{Arnett_1988}.  We neglect the photosphere, which includes only $\sim 1\%$ of the envelope mass and makes a negligible contribution to the scalar field profile.  

Figure~\ref{f:phiSN} shows $\phi(r)$ for several values of $\re$.  Note that $\phi$ changes along with the envelope until the gravitational potential $G \Me / \re \sim \Psis$, after which $\phi$ becomes linear in the envelope.  Linearity implies a conserved charge, hence no further change in the exterior field profile.  For the model shown, this occurs at $\re \approx 200\rsun$.  Although a more accurate treatment of the neutron star would include a more accurate equation of state~\cite{Babichev_Langlois_2010} and strong gravity effects~\cite{Upadhye_Hu_2009}, Fig.~\ref{f:phiSN} shows that the field in the envelope is insensitive to the details of the neutron star.

{\bf{\em Power radiated:}}
Consider $r \gg \tau$, and approximate the oscillation of $\re$ in a variable star as sinusoidal.  The time-dependent part of the field is then $\Delta \phi \, \sin(\omega r-\omega t)/2$, where $\omega = 2\pi/\tau$ and $\Delta\phi \equiv \phimax(r) - \phimin(r)$.  The corresponding energy density is $u = \frac{1}{2}(\frac{d\phi}{dt})^2 + \frac{1}{2}(\frac{d\phi}{dr})^2 \approx \omega^2 (\Delta\phi)^2 \cos(\omega r - \omega t )^2/4$.  
Averaging over a spherical shell of radius $r$ and thickness $\tau$, we find the power lost to monopole radiation,
\begin{eqnarray}
\Pmono 
&=& 
(r\Delta\phi)^2 \omega^2 \pi/2
\label{e:Pmono}
\approx 
(2 \beta \Mpl \Psis  \, \Delta r)^2 \omega^2\pi/2, 
\end{eqnarray}
where the approximation applies to the screened case.

A similar approach can be used to approximate the energy lost to monopole radiation during a supernova.  In our model, Fig.~\ref{f:phiSN}, $\phi(r,t)$ evolves with the expanding envelope until the envelope becomes linear at $\re \approx G \Me / \Psis$.  Since the inital value of $\re$ is much smaller, $\Delta r \approx G \Me / \Psis$.  This occurs in a time $\tau/2 \equiv \Delta r / \ve$, resulting in a total radiated energy of
\begin{equation}
\Emono
=
(r\Delta \phi)^2 \pi^3/\tau
\approx
\pi^2\beta^2 \Me \ve \Psis / 2.
\label{e:Emono}
\end{equation}

\section{Results and Discussion}
\label{sec:results_and_discussion}

\begin{figure}[tb]
\begin{center}
\includegraphics[angle=270,width=3.3in]{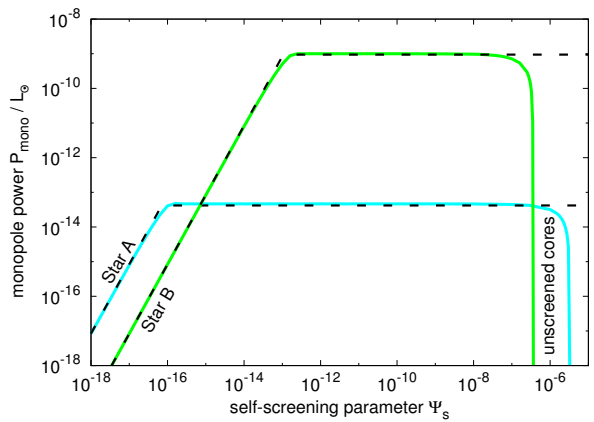}
\caption{Power emitted in monopole radiation for the variable stars in Table~\ref{t:stars}.  Solid lines use numerical computations for $\phi(r)$.  Dashed lines use the screened approximation of~(\ref{e:Pmono}) but cap $\Delta\phi/\Mpl$ at $\beta G M_\star \epsenv / {\bar r_E}$ to account for unscreened envelopes.  $\beta = 1/\sqrt{6}$ has been assumed.\label{f:Pmono}}
\end{center}
\end{figure}

{\bf{\em Energy loss constraints:}}
Although scalar monopole radiation cannot be detected directly, it may affect the dynamics of astrophysical objects in observable ways.  
Sufficiently powerful monopole radiation could quench the pulsation of a variable star.  The radiated power (\ref{e:Pmono}) increases with $\Psis$ as long as the envelope remains screened, so it will stop growing at $\Psis \sim G M_\star\epsenv/\re$.  This occurs at $\Psis \sim 10^{-16}$ and $10^{-13}$ for stars \calA~and \calB, respectively, resulting in bounds of 
\begin{equation}
P_{\mathcal{A}}
\lesssim
10^{-13}(\beta\sqrt{6})^2 \Lsun
\textrm{ and } 
P_{\mathcal{B}}
\lesssim
10^{-9} (\beta\sqrt{6})^2 \Lsun.
\label{e:Pmono_bounds}
\end{equation}
Figure~\ref{f:Pmono} shows the monopole power emitted by variable stars. These energy loss rates are too small to provide useful constraints unless the coupling strength $\beta$ is $5$-$6$ orders of magnitude larger than unity.

Next, consider supernova constraints.  The kinetic energy of the expanding envelope is  $\Ekin = 3\Me\ve^2/10$.  If~(\ref{e:Emono}) is close to this, then monopole radiation will significantly slow down the envelope and change its density profile.  We estimate constraints by requiring that
\begin{equation}
\frac{\Emono}{\Ekin}
=
\frac{5\pi^2}{3} \frac{\beta^2\Psis}{\ve}
=
8\times 10^{-5}\frac{\beta^2}{\frac{1}{6}}\frac{\Psis}{10^{-6}} \frac{10^4\frac{\mathrm{km}}{\mathrm{s}}}{\ve}
\label{e:E_ratio}
\end{equation}
be smaller than unity.  Note that this constraint is independent of the stellar mass and the details of the supernova, aside from the envelope velocity at a time $\Delta R /\ve \sim 1-10$~hours after core collapse.  

For currently allowed models of $f(R)$ gravity~\cite{Jain_Vikram_Sakstein_2012}, $\beta=1/\sqrt{6}$ and $\Psis \lesssim 10^{-6}$, monopole constraints $\Psis \lesssim 10^{-2}$ are not competitive.  
For generalized modified gravity models in which $\beta$ can be arbitrary, monopole radiation constraints are more powerful.  Screening in our galaxy still requires $\Psis \lesssim 10^{-6}$ in such models.  For $\Psis = 10^{-6}$ (\ref{e:E_ratio}) requires $\beta \lesssim 100$.

{\bf{\em Environmental screening:}}
Our previous calculations assume that the variable star or supernova is in an unscreened region, so that $\phi \rightarrow \phiB(\rhomo) = 2\beta\Mpl\Psis$ far from the star.  If the star is near a galaxy or galaxy cluster, then the field will instead approach some lower value $2\beta\Mpl\Psie$, where $\Psie$ is determined by the environment.  For example, a screened galaxy in a large cluster can easily have a matter overdensity of $\rho/\rhomo-1 \sim 10^5$, which for $n=1/2$ implies $\Psie \sim 10^{-10} \Psis$.  

For a variable star, monopole power is largest where the core is screened but the envelope is not.  Environmental screening may move the star into or out of this region, but the upper  bound~(\ref{e:Pmono_bounds}) on $\Pmono$ still applies.  Thus the energy loss through monopole radiation from variable stars does not provide a significant constraint on modified gravity.    

In a supernova, environmental screening weakens the constraint~(\ref{e:E_ratio}), replacing $\Psis$ by $\Psie \leq \Psis$.  For $\Psis \sim 10^{-6}$, at the level of approximation used here, environmental screening should be negligible for a supernova in a dwarf galaxy, such as SN1987a.  Thus $\Psie \sim \Psis$ and our constraint from~(\ref{e:E_ratio}), $\beta \lesssim 100$, should still be valid.
More precise constraints may be found using a screening map~\cite{Cabre_etal_2012}.  Through such a map it may be possible to find unscreened halo supernovae~\cite{Sand_etal_2011} for $\Psis$ as low as $10^{-8}$.  

{\bf{\em Conclusions:}} 
We have estimated the energy loss due to scalar monopole radiation in supernovae and variable stars, in $f(R)$ gravity and its generalization to arbitrary matter couplings $\beta$.  For standard $f(R)$ gravity, we conclude that this energy loss is too small to provide competitive constraints, or to affect the stellar luminosity and variability calculations of~\cite{Davis_etal_2012,Jain_Vikram_Sakstein_2012}.  In particular, energy conservation during stellar pulsation implies that the oscillating stellar envelope contains only a small fraction of the total stellar mass, limiting the power emitted as monopole radiation.

Constraints on generalized $f(R)$ gravity from monopole radiation are more useful.  In order to estimate them we used numerical solutions of the scalar field equation for static matter configurations corresponding to the endpoints of stellar oscillation.  In the far-field regime, the scalar perturbation from its background value decreases as $1/r$, so differences due to stellar pulsation persist far from the star.  Using these differences we have computed the energy density and power loss in monopole radiation.  For both dwarf and classical Cepheid variables we find that the radiated power is, at most, $\sim 10^{-12}\beta^2$ times the luminosity of the star.  Thus $\beta \sim 10^5 - 10^6$ is consistent with Cepheid variables.

Core-collapse supernovae provide more powerful constraints, since the amount of matter ejected during a supernova explosion is much greater than that in a stellar envelope.  We model this ejected matter as a uniform-density sphere undergoing homologous expansion.  Requiring that the energy lost to monopole radiation be small compared with the total kinetic energy of the expanding envelope, we find~(\ref{e:E_ratio}).  Though a precise constraint on $\beta$ and $\Psis$ would require a screening map~\cite{Cabre_etal_2012}, we estimate that SN1987a is unscreened for $\Psis \gtrsim 10^{-6}$, yielding the constraint $\beta \lesssim 100 (\Psis/10^{-6})$.  

{{\bf{\em{Acknowledgments:}}}
We are grateful to K.~Barbary, S.~Habib, C.~Hirata, W.~Hu, B.~Jain, S.~Kuhlmann, A.~Nelson, J.~Sakstein, F.~Schmidt, and V.~Vikram for insightful discussions. 
A.U.~was supported by the U.S. Department of Energy, Basic Energy Sciences, Office of Science, under contract No.~DE-AC02-06CH11357. 
J.H.S. is supported through the Center for
Interdisciplinary Exploration and Research in Astrophysics (CIERA) at
Northwestern University.

The submitted manuscript has been created by
UChicago Argonne, LLC, Operator of Argonne
National Laboratory (``Argonne''). Argonne, a
U.S. Department of Energy Office of Science laboratory,
is operated under Contract No. DE-AC02-
06CH11357. The U.S. Government retains for itself,
and others acting on its behalf, a paid-up
nonexclusive, irrevocable worldwide license in said
article to reproduce, prepare derivative works, distribute
copies to the public, and perform publicly
and display publicly, by or on behalf of the Government.

\bibliographystyle{unsrt}
\bibliography{chameleon}
\end{document}